\documentclass[12pt]{iopart}
\usepackage{graphicx}
\usepackage{epsfig}
\newcommand{\mubold}{\mbox{\boldmath{$\mu$}}}
\begin{document}
\title{A muon-spin relaxation study of BiMnO$_{3}$}
\author{T. Lancaster$^{1}$, S.J. Blundell$^{1}$, P.J. Baker$^{1}$,
F.L. Pratt$^{2}$, W. Hayes$^{1}$, I. Yamada$^{3}$, M. Azuma$^{4}$ 
and M. Takano$^{4}$}
\ead{t.lancaster1@physics.ox.ac.uk}
\address{$^{1}$Clarendon Laboratory, Oxford University Department of
  Physics, Parks Road, Oxford, OX1 3PU, UK}
\address{$^{2}$ISIS Facility, Rutherford Appleton Laboratory, Chilton,
  Oxfordshire OX11 0QX, UK}
\address{$^{3}$ Institut de Min\'{e}ralogie et de Physique des Milieux 
Condens\'{e}s Universit\'{e} Pierre et Marie Curie, 75015 Paris, France}
\address{$^{4}$Institute for
Chemical Research, Kyoto University, Uji, Kyoto 611-0011, Japan}

\begin{abstract}
We present the results of muon-spin relaxation measurements on
ferromagnetic BiMnO$_{3}$.
Below
$T_{\mathrm{C}}=98.0(1)$~K oscillations in the
time-dependence of the muon polarization are observed, characteristic of a quasistatic magnetic field
at a single muon site, allowing us to probe the critical behaviour
associated with the magnetic phase transition. 
We are able to suggest candidate muon sites on the basis of dipole field 
calculations.
Close to $T_{\mathrm{C}}$, fluctuations of the Mn$^{3+}$ moments are
characteristic of critical behaviour while there is a sharp crossover to
a region of fast dynamic fluctuations at higher temperatures. 
\end{abstract}

\submitto{\JPCM}
\vspace{1cm}
\begin{flushright}
(\today)
\end{flushright}
 
\maketitle
\section{Introduction}

The possibility of combining ferroelectric and magnetic order in a
material is of great interest from the point of view of fundamental
condensed matter physics and of technology \cite{fiebig}. Although the microscopic nature of
materials usually excludes this coexistence, there are a growing
number of systems where spontaneous electrical and magnetic polarization 
exist in the same phase. 
These are examples of multiferroic materials (the general term
describing systems where more than one of the ferroic 
order parameters are nonzero in the same phase)
and have been of growing interest
in recent years \cite{cheong,khomskii,eerenstein}. The system $R$MnO$_{3}$ 
(where $R$ is a trivalent ion) has been widely studied in this
context due to the coexistence, in
some of the members of this series, of ferroelectric and magnetic
order. The nature of the $R^{3+}$ ion is central to determining the
structural, ferroelectric and magnetic properties of members of this system. 

\begin{figure}
\begin{center}
\epsfig{file=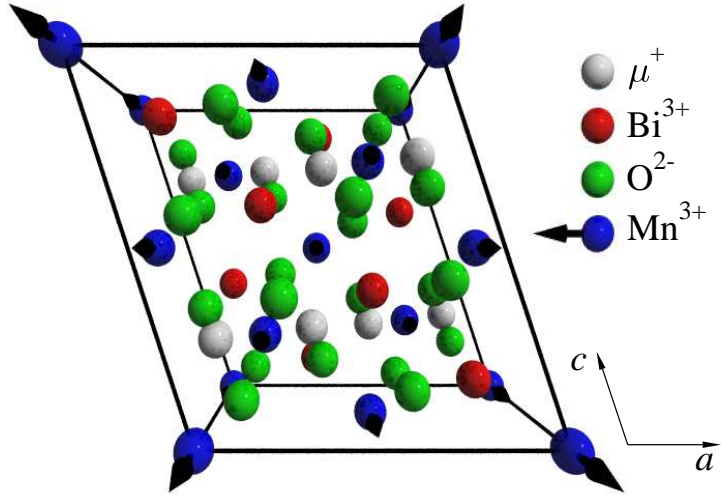,width=10cm}
\caption{Structure of BiMnO$_{3}$ ($C2$ space group) viewed down the $b$-axis.
Bi is shown in red, Mn in blue, O in green and the
proposed muon stopping sites in white (see main text).
Arrows show the ferromagnetic spin structure found below
$T_{\mathrm{C}}$, with the moments all along the b-axis. \label{muonsite}}
\end{center}
\end{figure}

BiMnO$_{3}$ has been the subject of much research interest, partly 
because its relative structural simplicity
makes it amenable to study using first-principals theoretical 
techniques \cite{hill}. 
Early experimental work on the material was sparse, in part due to the
requirement of high pressures in the preparation of the compound, and
reports of ferroelectricity were initially
speculative \cite{early}. A detailed structural study
\cite{atou} suggested 
that the material has a highly distorted perovskite
structure (monoclinic space group $C2$), shown in figure
(\ref{muonsite}), 
which is consistent with the existence of ferroelectricity. 
Subsequent theoretical \cite{hill} and experimental 
studies \cite{moreira_ssc,moreira,kimura,chi} also pointed to a
coexistence of ferromagnetic and ferroelectric ordering at low temperature. 
Ferroelectric ordering was proposed to take place below 
$T_{\mathrm{E}}\approx 770$~K \cite{kimura}
induced by a centrosymmetric to non-centrosymmetric structural
transition, although it
was also suggested \cite{moreira_ssc} that the ferroelectricity
accompanied a structural transition
occurring at 450~K without a change in symmetry.

The different structure and magnetic properties 
of BiMnO$_{3}$ in comparison to other $R$MnO$_{3}$ manganites, 
(which are antiferromagnetic and either orthorhombic or hexagonal)
has been ascribed to 
the presence of the lone pair on bismuth \cite{seshadri,moreira} and 
the covalent nature of the Bi--O bonds which introduce
additional orbital interactions (in contrast to the other $R$MnO$_{3}$
manganites, where the bonding is essentially  ionic
\cite{hill}). 
It was proposed that these interactions lead to orbital order below the
centrosymmetric to non-centrosymmetric transition at 770~K
\cite{yang}.
The orbital order, in turn, fosters three dimensional ferromagnetic 
interactions
in this material \cite{moreira} causing the material to undergo a 
ferromagnetic ordering transition 
at a Curie temperature $T_{\mathrm{C}}$, reported to be
$T_{\mathrm{C}} = 98$--$105$~K \cite{sugawara,chiba,moreira_ssc,belik}.
The magnetic ordering transition results in a collinear magnetic
structure \cite{moreira} where the 
Mn$^{3+}$ ions (3d$^{4}$, $S=2$) align along the $b$ axis 
(figure (\ref{muonsite})). 

Recently, detailed stuctural studies have suggested
that, in contrast to the state of affairs described above,
BiMnO$_{3}$ actually crystallizes in the centrosymmetric space group
$C2/c$ at 300~K \cite{belik,belik2,montanari2}. This is incompatible
with the existence of ferroelectricity and may necessitate a major
revision of our view of this material. For example, it has been
suggested that the large
dielectric anomaly observed near $T_{\mathrm{C}}$ may be linked to
structural distortions occurring near this temperature 
regime rather than to effects related to broken inversion symmetry 
\cite{montanari2}. 

He we confine our discussion to the magnetic properties
of BiMnO$_{3}$.
Muons are a sensitive probe of magnetism in condensed matter \cite{steve}
and allow us to investigate magnetism from a local viewpoint. 
In this paper we present the results of a zero-field (ZF)
muon-spin relaxation ($\mu^{+}$SR) study of the local magnetic
properties of BiMnO$_{3}$. We have
probed the magnetic field distribution both below and above 
$T_{\mathrm{C}}$, allowing us to follow the temperature evolution of
the order parameter in the magnetically ordered phase and to investigate
the magnetic fluctuations in the paramagnetic phase. Our measurements
also allow us to identify candidate muon stopping sites in the material. 

\section{Experimental details}

A polycrystalline sample of BiMnO$_{3}$ was prepared 
using the method reported previously \cite{kimura}, which involves 
heating a mixed powder of Bi$_{2}$O$_{3}$, Mn$_{2}$O$_{3}$ and
MnO$_{2}$ to 700~$^{\circ}$C under 3~GPa of pressure for 30 minutes.
The structural character of the sample was checked using x-ray
diffraction, while the bulk magnetic properties of the sample 
were measured using a commercial magnetometer.

In a $\mu^{+}$SR experiment, spin-polarized
positive muons are stopped in a target sample, where the muon usually
occupies an interstitial position in the crystal.
The observed property in the experiment is the time evolution of the
muon spin polarization, the behavior of which depends on the
local magnetic field $B$ at
the muon site, and which is proportional to the
positron asymmetry function  $A(t)$ \cite{steve}.
ZF $\mu^{+}$SR  measurements were made 
on  BiMnO$_{3}$ 
using the GPS instrument at the Swiss Muon Source (S$\mu$S), 
Paul Scherrer Institute, Villigen, Switzerland.
The sample was wrapped in a silver foil packet (foil
thickness 25~$\mu$m) and mounted on a silver backing plate. 
Silver is used as it possesses only a small
nuclear moment and so minimizes any background depolarizing signal.

\section{Results}

\begin{figure}
\begin{center}
\epsfig{file=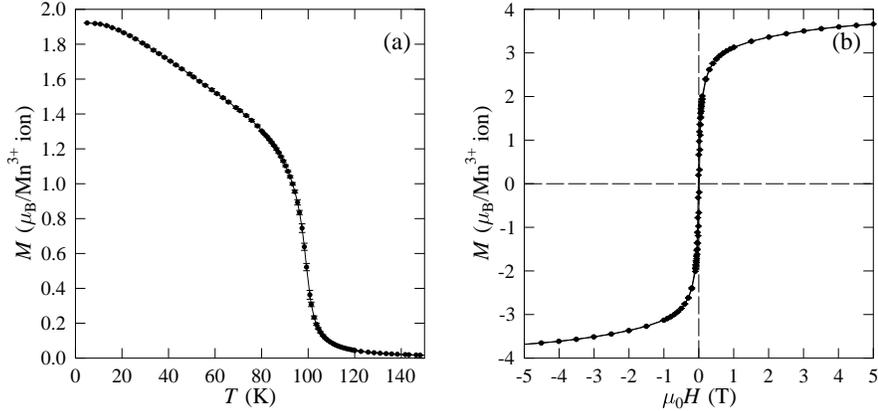,width=12cm}
\caption{Magnetization of BiMnO$_{3}$:
(a) measured as a function of temperature in a field of 100~mT;
(b) measured as a function of applied field at $T=5$~K.
\label{magnetization}
}
\end{center}
\end{figure}

Figure \ref{magnetization} shows the results of magnetization measurements made on our sample
as a function of temperature and applied magnetic field. As seen in
previous studies \cite{sugawara,chi,moreira_ssc,kimura,belik}, the magnetic
 transition is evident close to 100~K (figure \ref{magnetization}(a)). In an applied magnetic
field the magnetic moment is seen, at low temperatures,  to approach a saturation value of
around $\sim 3.8\mu_{\mathrm{B}}$ (figure \ref{magnetization}(b)), 
as was also observed
previously (see above). 

\begin{figure}
\begin{center}
\epsfig{file=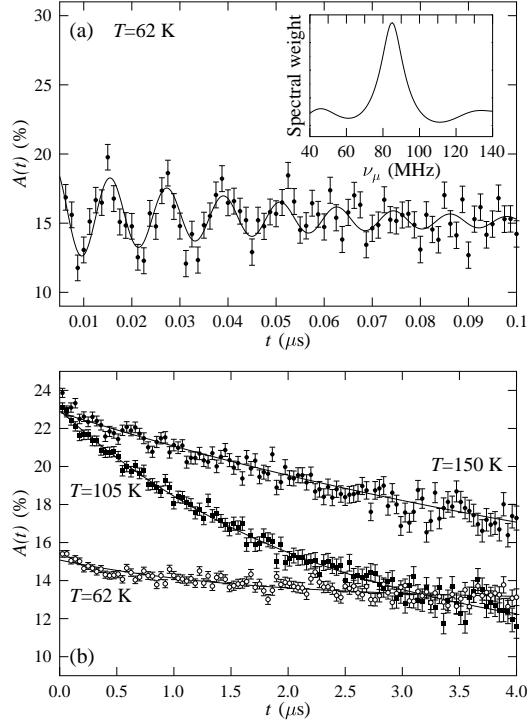,width=7cm}
\caption{ZF $\mu^{+}$SR spectra measured for
BiMnO$_{3}$, showing fits as described in the text. 
(a) Example spectrum for times $0 \leq t \leq 0.1~\mu$s
measured below $T_{\mathrm{C}}=98.0$~K. Oscillations are observed which are 
characteristic
of quasistatic magnetic order at the muon site.
{\it Inset:} Maximum entropy spectrum showing a single muon-spin
precession frequency.
(b) Spectra for times $0 \leq t \leq 4~\mu$s. 
Above $T_{\mathrm{C}}$
the spectra are well described by a single exponential relaxation
function. Below $T_{\mathrm{C}}$ only the longitudinal spin component
is resolvable, with relaxation due to dynamic fluctuations.
\label{bidat}}
\end{center}
\end{figure}

Example ZF $\mu^{+}$SR spectra measured at S$\mu$S on BiMnO$_{3}$ are
shown in figure~\ref{bidat}. 
At temperatures below $T=100$~K and for times $t<0.2~\mu$s,
we observe oscillations in the asymmetry spectra (figure~\ref{bidat}(a)). 
These are characteristic of a quasistatic local
magnetic field at the muon site, which causes a coherent precession of 
the spins of those muons with a component of their spin 
polarization perpendicular to this local field; their presence
provides strong evidence for the existence of long range magnetic order
in this phase, in agreement with previous magnetic measurements.
The frequency of the oscillations is given by 
$\nu_{i}=\gamma_{\mu} B_{i}/2 \pi$, where
$\gamma_{\mu}$ is the muon gyromagnetic ratio 
($\equiv 2 \pi \times 135.5$~MHz T$^{-1}$) and $B_{i}$ is the local field at
the $i$th muon site. In the presence of a distribution of
local magnetic fields the oscillations are expected to relax with
a depolarization rate $\lambda$. One precession frequency is observed
in the measured spectra (see {\it inset} figure~\ref{bidat}(a)), 
corresponding to a magnetically unique muon site
in the material.

The spectra  in this temperature regime  were found to be
best fitted, for times $t<1~\mu$s, to a function of the form
\begin{equation}
A(t)= A_{\mathrm{bg}}+ A_{\mathrm{rel}} \left( \frac{1}{3}\exp(-\lambda_{1} t)
+ \frac{2}{3} \exp(-\lambda_{2} t) \cos (2 \pi \nu_{\mu} t
+ \phi) \right), \label{fitbi}
\end{equation}
where $A_{\mathrm{rel}}$ represents the signal from the sample
and 
$A_{\mathrm{bg}}$ represents a constant background contribution from those
muons that stop in the sample holder or cryostat tail.
The term multiplied by 1/3 accounts for those components
of the muon spin that are expected to lie parallel to the local magnetic
field at the muon site in a polycrystalline material (and would therefore not be expected to 
give rise to an oscillatory signal). 
The transverse relaxation rate $\lambda_{2}$ was found to be constant
for each temperature $T < T_{\mathrm{C}}$ and was fixed at a value
$\lambda_{2}=7.0$~MHz.
A phase offset of $\phi=-115^{\circ}$ was also required to fit the
oscillations, probably arising from the difficulty in exactly 
identifying $t=0$ in this measurement of a very fast precession signal.
We note that the model that gives rise to equation (\ref{fitbi}) describes
a material that is uniformly magnetically ordered throughout its
bulk. 
In our $\mu^{+}$SR measurements we are unable to resolve any signal
from a spin-glass-like phase that has recently been proposed
to exist below $T_{\mathrm{C}}$ \cite{belik}.

\begin{figure}
\begin{center}
\epsfig{file=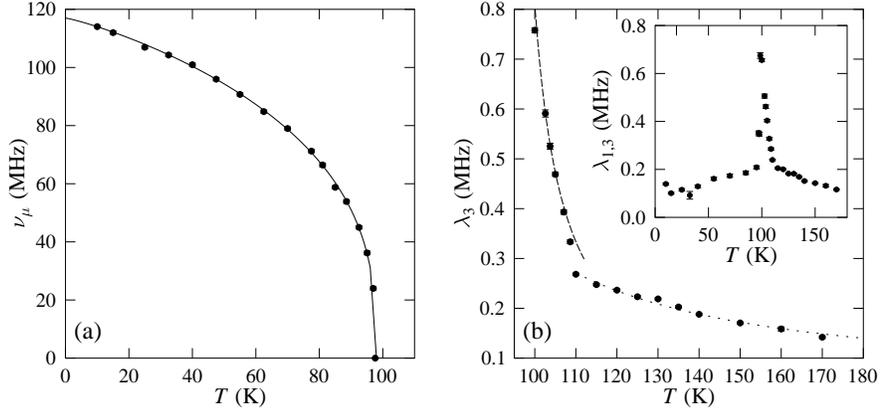,width=12cm}
\caption{
(a) Temperature evolution of the muon precession
frequency below $T_{\mathrm{C}}=98.0$~K with a fit to the
phenomenological form
$\nu_{\mu}=\nu_{\mu}(0)(1-(T/T_{\mathrm{C}})^{\alpha})^{\beta}$.
(b) Variation of the relaxation rate $\lambda_{3}$ for
spectra measured above $T_{\mathrm{C}}$. A large
change in slope is seen close to 110~K. A Fit is shown to
a model of critical dynamics  for $T <T_{\mathrm{cr}}$ (dashed line;
see text).
The dotted line for $T >T_{\mathrm{cr}}$ is a guide to the eye.
{\it Inset:} Dynamic relaxation across the entire measured
temperature range 
($\lambda_{1}$ for $T < T_{\mathrm{C}}$, $\lambda_{3}$ for $T > T_{\mathrm{C}}$)
showing that the crossover is roughly symmetric
about $T_{\mathrm{C}}$ (see main text). 
\label{bian}}
\end{center}
\end{figure}

The temperature evolution of the frequency,
extracted from fits to equation~(\ref{fitbi}) to the
measured data, is shown in figure~\ref{bian}(a).
Fits to the phenomenological form 
$\nu_{\mu}=\nu_{\mu}(0)(1-(T/T_{\mathrm{C}})^{\alpha})^{\beta}$
 yield $T_{\mathrm{C}}=98.0(1)$~K, $\nu_{\mu}(0)=117(1)$~MHz,
$\alpha=1.1(1)$ and $\beta=0.35(1)$. 
We note that our value of $\beta$ is consistent with the 3D Heisenberg
model. The parameter $\alpha$ is smaller than would be expected 
for ferromagnetic magnons (for which we expect $\alpha=3/2$).
These parameters are robust when the fitting range is varied
and attempts to fit only the low temperature portion of the data similarly
fail to reveal the expected low temperature dependence. Although this
may be attributed to the low concentration of data points in the
limit $T \rightarrow 0$, we note that the form for the
magnetic contribution to the heat capacity ($C_{\mathrm{m}} \propto T^{3/2}$)
expected for ferromagnetic magnons was not observed \cite{belik},
which might suggest that the system is not amenable to a simple spin wave
analysis.

Our estimate of $\nu_{i}(0)$ allows us to attempt to determine the muon
sites in BiMnO$_{3}$. The local magnetic field at a muon site 
{\bf B}$_\mu$({\bf r}$_{\mu}$) is usually dominated by
the sum of dipolar contributions. Assuming the local field is due
entirely to the dipole field of Mn$^{3+}$ ions we may write
\begin{equation}
\mathbf{B}_{\mu}(\mathbf{r}_{\mu}) = \sum_{i} \frac{3(\mubold_{i}
\cdot \hat{\mathbf{n}}_{i})\hat{\mathbf{n}}_{i} - \mubold_{i}}{\mid
\mathbf{r}_{\mu} - \mathbf{r}_{i} \mid ^{3}},
\end{equation}
where {\bf r}$_{\mu}$ is the position of the muon, $\mubold_{i}$ is the
ordered magnetic moment of the $i^{\mathrm{th}}$ Mn ion and $\hat{{\mathbf
n}}_{i}$ ($= (\mathbf{r}_{\mu}-  \mathbf{r}_{i}) / \mid
\mathbf{r}_{\mu} - \mathbf{r}_{i} \mid$)
is the unit vector
from the muon to the Mn ion at site {\bf r}$_{i}$. The positive muon's
position is usually in the vicinity of the electronegative O$^{2-}$ 
ions \cite{holzschuh}. 

Dipole fields were calculated in a sphere containing $\approx 10^{5}$
Mn ions with moments of 3.8$\mu_{\mathrm{B}}$ (that is, the value of
the saturation magnetization per Mn$^{3+}$ ion estimated from our
magnetization measurements) 
aligned along the [010]
direction \cite{moreira}.  
A magnetic field consistent with $\nu_{i}(0)$ is found at several positions in the unit
cell. Using the $C2$ space group, candidate muon sites may be
identified at the positions (0.40, 0.52, 0.26) and 
 (0.10, 0.67, 0.24), placing muons either side of the Mn$^{3+}$ ion
located at 
(0.246 0.563 0.243).
Application of the symmetry elements of $C2$ generates 
crystallographically equivalent sites 
at which the same local field is experienced. 
The result is that the magnetically equivalent candidate muon
sites lie close
to a Mn$^{3+}$ ion near $z=0.25$ or $z=0.75$, as shown in figure
\ref{muonsite}, where 8 sites are evident per unit cell.
Applying the elements of $C2$ to either candidate site generates no
positions at which a different 
local magnetic field is found, which
is consistent with our observation that 
there is a unique local field at all muon sites in the
ordered regime.
The calculation was checked for several values
of Mn$^{3+}$ magnetic moment in the range 3.2$\mu_{\mathrm{B}} \leq \mu_{i} \leq
3.8 \mu_{\mathrm{B}}$ to account for the different values of the
magnetic moment reported from previous work 
\cite{moreira_ssc,moreira,kimura,chi,belik}. Very similar
 muon sites to those described above were found for all calculations, 
with the exact positions shifted by a maximum of $\sim 2$\% of the
lattice parameters. 
Repeating this procedure using the space group $C2/c$ leads to little
difference in the local magnetic field profile. In this case the muon
position is
given by (0.60, 0.71, 0.50).
Applying the symmetry elements of
$C2/c$ gives similar muon sites to those described above, 
placing muons on either side of the Mn ions (0.75, 0.75, 0.5)
and again leading to
8 magnetically equivalent sites per unit cell.
It should be noted that in a ferromagnet such as 
BiMnO$_{3}$ we would also expect contributions from the Lorentz 
field, demagnetization field and the hyperfine contact field which 
may mean that the situation is somewhat more complicated than considered here.

Above $T_{\mathrm{c}}$, the spectra are found to be purely relaxing 
 (figure~\ref{bidat}(b)) and are best described with a
single exponential function
\begin{equation}
A(t) = A_{\mathrm{bg}}+ A_{\mathrm{rel}} \exp (- \lambda_{3} t),\label{exp}
\end{equation}
where $\lambda_{3}$ is a relaxation rate and $A_{\mathrm{rel}}$ is
constant. An exponential relaxation is usually indicative of dynamic
fluctuations in the local magnetic field at the
muon site \cite{hayano}. In this case the relaxation rate may be
written as $\lambda_{3} \propto (\Delta B)^{2} \tau$, where $\Delta B$ is
the second moment of the local magnetic field distribution and $\tau$ 
is the correlation time. If the distribution of spin disorder 
within the system 
is temperature
dependent we would expect a variation in $\Delta B$. More often,
in the absence of magnetic or structural transitions,  
$\Delta B$ does not have a dramatic temperature variation and it is
the fluctuation rate $\tau$ that varies with temperature. 
The variation of $\lambda_{3}$ with temperature 
is shown in figure~\ref{bian}(b) where we see a monotonic decrease in
$\lambda_{3}$ with increasing temperature but with a 
significant change in
slope around $T_{\mathrm{cr}}\approx 110$~K. 
This is indicative of a sharp crossover between regimes of
slow dynamics (large $\tau$) below $T_{\mathrm{cr}}$ and faster
dynamics (small $\tau$) at higher temperatures.  
A clue to the origin of this behaviour may be found by comparing
$\lambda_{3}$ to the relaxation at long times ($t<10$~$\mu$s)  for $T <
T_{\mathrm{C}}$, corresponding to the relaxation rate
$\lambda_{1}$ in equation~(\ref{fitbi}), which also arises due to
dynamic relaxation processes. 
This is facilitated by increasing the size of the time bins of the
data so that the oscillations become unresolvable below
$T_{\mathrm{C}}$ as shown in figure \ref{bidat}(b). 
The results of such an analysis are shown in the inset of
figure~\ref{bian}(b), where we see that
$\lambda$ is roughly symmetrical in temperature about
$T_{\mathrm{C}}$, 
with $\tau$ increasing dramatically for $|T-T_{\mathrm{C}}| < 10$~K. 

The sharp crossover seen in the dynamic relaxation rates 
$\lambda_{1}$ and $\lambda_{3}$ does not appear to correspond to
features in the the bulk properties of BiMnO$_{3}$. Given the
symmetrical nature of the effect about the critical temperature, 
it seems unlikely that it is
related to a structural transition (causing a change in the
second moment of the local field distribution $\Delta$) of the sort that has 
been suggested to account for the dielectric anomaly
 near $T_{\mathrm{C}}$ \cite{montanari2}. Instead, it is probable that
it is a different dynamic relaxation channel that determines the
relaxation time $\tau$ on either side of the crossover.
In a material where more than one relaxation channel coexists at
a particular temperature, only the {\it slowest} (i.e.\ the
channel described by the largest relaxation rate $\tau$) 
will dominate the muon-spin
relaxation. This is expected to lead to relaxation with a single
relaxation rate $\lambda$ as observed in this case. (In contrast,
multiple 
relaxation rates only coexist in cases
where the fluctuations are confined to spatially separate regions 
\cite{heffner}.) 
From our observation of a single
relaxation rate, we may therefore infer that the fluctuations 
occur throughout the entire bulk of BiMnO$_{3}$ at all temperatures.
Furthermore, the fact
that the amplitude $A_{\mathrm{rel}}$ is constant across the measured
temperature range shows that this
crossover is intrinsic to the bulk of the material.

At temperatures above $T_{\mathrm{cr}}$ dynamic fluctuations of the
Mn$^{3+}$ spins are probable, slowing as the temperature is decreased leading
to an increase in $\lambda_{3}$. 
The nature of the dynamics causing the muon depolarization
in this regime is not clear and its temperature dependence
does not enjoy a form that may be described by a simple physical model. 
As the temperature is decreased
below $T_{\mathrm{C}}$ another relaxation process, with a more dramatic
temperature dependence, becomes the slowest channel and then dominates
the measured relaxation rate $\lambda$.
The symmetrical behaviour of $\lambda_{1,3}$ 
suggests that this process is caused by 
critical dynamics \cite{hohenberg} in the region 
$(T_{\mathrm{C}} - \Delta) \leq T \leq (T_{\mathrm{C}}+\Delta$), 
where $\Delta = T_{\mathrm{cr}}-T_{\mathrm{C}}$.
Dynamic relaxation rates are proportional to 
$\sum_{\mathbf{k}}C(\mathbf{k},\omega=0)$, where
$C(\mathbf{k},\omega=0)$
is the zero frequency correlation function for wavevector {\bf k},
describing the local fields at the muon site \cite{hohenberg}. 
Calculations of $C(\mathbf{k},\omega=0)$ lead us to expect $\lambda_{3}$ to
diverge close to $T_{\mathrm{C}}$ according to 
$\lambda_{3} \propto |T-T_{\mathrm{C}}|^{-w}$, where $w$ may be related
to the standard critical exponents \cite{binney} 
via $w= \nu(z+2-\eta-d)$ \cite{hohenberg}. For a $d=3$ Heisenberg
model we find $w=1.026$ \cite{pelisseto}, leading to the fit
shown in figure~\ref{bian}(b) for $T_{\mathrm{C}} \leq T \leq
T_{\mathrm{cr}}$. This approach describes the data adequately, becoming
worse, as should be expected, as we move away from $T_{\mathrm{C}}$.
Finally, as the temperature is lowered below $T_{\mathrm{C}} - \Delta$
in the ordered regime, residual spin fluctuations become the
slowest channel and therefore dominate $\lambda_{1}$. 

These results suggest that the spin fluctuations in BiMnO$_{3}$
are not trivial.
Inelastic neutron scattering measurements may throw further light on
the magnetic excitations in this system.


\section{Conclusions}

We have carried out a $\mu^{+}$SR investigation of the multiferroic material
BiMnO$_{3}$. Our measurements show that the
ferromagnetic transition occurs at $T_{\mathrm{C}}=98.0(1)$~K with two
regimes of dynamics separated by a sharp crossover at 
$T_{\mathrm{cr}} \approx 110$~K. We have also identified
candidate muon sites in the material on the strength of dipole field calculations. 
Muons have been shown to be a useful and effective probe of magnetism in this system.
The possibility of carrying out future muon studies in the presence of applied electric 
fields (for an example see \cite{storchak}) presents opportunities for 
investigations
of the microscopic mechanisms responsible for multiferroic behaviour in this
class of materials. 

\ack
Part of this work was carried out at the Swiss Muon Source, 
Paul Scherrer Institute, Villigen, Switzerland. We thank Alex Amato
 for technical assistance. This work is supported by the EPSRC. 
T.L.\ acknowledges support from the Royal Commission for the Exhibition
of 1851.

\end{document}